# Archer: A Community Distributed Computing Infrastructure for Computer Architecture Research and Education

Renato Figueiredo, P. Oscar Boykin, José A. B. Fortes, Tao Li,
Jie-Kwon. Peir, David Wolinsky (University of Florida)
Lizy John (University of Texas at Austin)
David Kaeli (Northeastern University)
David Lilja (University of Minnesota)
Sally McKee (Cornell University)
Gokhan Memik (Northwestern University)
Alain Roy (University of Wisconsin-Madison)
Gary Tyson (Florida State University)

**Abstract**
This paper introduces Archer, a community-based computing resource for computer architecture research and education. The Archer infrastructure integrates virtualization and batch scheduling middleware to deliver high-throughput computing resources aggregated from resources distributed across wide-area networks and owned by different participating entities in a seamless manner. The paper discusses the motivations leading to the design of Archer, describes its core middleware components, and presents an analysis of the functionality and performance of a prototype wide-area deployment running a representative computer architecture simulation workload.

## 1. Introduction

Modern computer architecture research is driven by quantitative analysis. Leading-edge research requires detailed, cycle-accurate evaluation of many benchmark applications with several simulated configurations and is thus tightly dependent on the availability of high-throughput computing (HTC) systems. Many research groups are hindered in their ability to perform research because of lack of access to such resources. This is because, in addition to hardware costs, the investment of time and funds to train and educate students and staff to deploy, maintain and effectively use such systems presents a significant barrier of entry, especially for small- to medium-sized research groups. This paper describes *Archer*[1], a community-based computing resource for computer architecture research and education. Archer integrates technologies for resource virtualization, batch job schedulers, and multi-institution collaboration, in order to create:

- <u>A computing infrastructure which scales in capacity with community buy-in</u>: Archer starts from a seed set of cluster resources deployed at the Florida Statue University, Northeaster University, University of Texas at Austin, Northwestern University, University of Minnesota, Cornell University, and University of Florida. Subsequently, each new user joining Archer with one or more desktops or servers seamlessly contribute to its aggregate capacity.

- <u>A system that is easy for non-experts to join and use</u>: Archer relies on packaging and distribution of software environments for HTC as self-configuring virtual networks of virtual appliances, which can easily be installed by individual users in their own resources. Surveys from users of the virtual appliance used as a basis for Archer shows that users with no prior experience can typically install and use the system within 30 minutes.

---

[1] The Archer community infrastructure and Wiki are accessible at: http://archer-project.org

- <u>A community-based repository of simulation environments</u>: Archer allows sharing not only of hardware resources, but also of full-fledged software simulation modules consisting of application executables, support scripts, input and output data sets, and usage documents. In doing so, Archer facilitates the dissemination of useful tools and data sets, and foster creation of reproducible simulation experiments.

The community-driven features in Archer provide a new way to swiftly create grids of medium size, differentiating it from related infrastructures such as the Open Science Grid (OSG) and TeraGrid, in three important ways. First, Archer enables *seamless addition of resources* by the community, at a fine grain (at a minimum a single desktop computer by an individual user), within minutes. This is in contrast to OSG and TeraGrid, where individual resources cannot be easily incorporated, and to gain access to resources often takes days or weeks. Second, Archer deployments are virtualized and can be *easily replicated,* both at a smaller scale within an institution, and at a multi-institution scale by research communities. Archer's replicability enables research groups to easily bring up local Archer pools and be assured of preemptive access to their resources when needed, while providing opportunistic cycles to the community. This is in contrast to OSG and TeraGrid, which are large-scale shared physical resources not easily replicable at a small scale on local resources. Third, Archer empowers entry-level users to quickly learn HTC skills, from basic to advanced, with a combination of examples tailored to computer architecture and an interactiveinterface hosted on their own workstations. This is in contrast to OSG and TeraGrid, where entry-level users need to learn how to operate resources that are hosted remotely, using non-interactive sessions and unfamiliar interfaces for data transfer, login, and job scheduling. Figure 1 presents an overview of the Archer infrastructure.

## 2 Background and motivations

In modern computer architectures, processor performance, power consumption and cost are significantly affected by design parameters and target workloads. Thus, researchers rely on simulation environments to evaluate the merit of a new idea before it is implemented in hardware. In addition, computer architects depend on high-fidelity, cycle-accurate simulation environments (including the simulators themselves and associated tools such as compilers and datasets). Because these are complex and time-consuming to develop, researchers have relied on open-source extensible simulation environments, benchmarks and datasets developed by others in the community – such as SimpleScalar [5], SESC [29], PTLsim, RSIM, Wisconsin WARTS, among others – as well as on open-source modules that plug in to commercial systems, such as GEMS for Simics.

Computer architecture researchers' broad need to access high-performance resources and share simulation environments are addressed in an integrated manner by Archer. We believe that the availability of Archer encourages collaboration among groups by greatly simplifying the dissemination of applications, and increases the competitiveness of smaller research groups by providing seamless access to hardware resources and software environments. To illustrate use cases and the unique capabilities enabled by Archer, consider the scenarios described in Table 1 and illustrated by the following three fictional examples:

*Scenario 1: High-throughput cycles for research:* Graduate student Maria at Florida State U. is preparing a paper on a novel cache design for submission to a conference. She has developed a simulator which models her design. Each simulation takes on average 12 hours to complete on her desktop, and she wishes to analyze 10 configurations on 16 SPEC CPU benchmarks. The time to run this experiment on her desktop is prohibitively large (80 days). She downloads, instantiates an Archer appliance, copies her Linux simulator binary to the VM, prepares a Condor job file (building on a tutorial), and queues 160 jobs. Archer resources are utilized at 75% capacity by other jobs; still, her simulations are expected to finish within a day.

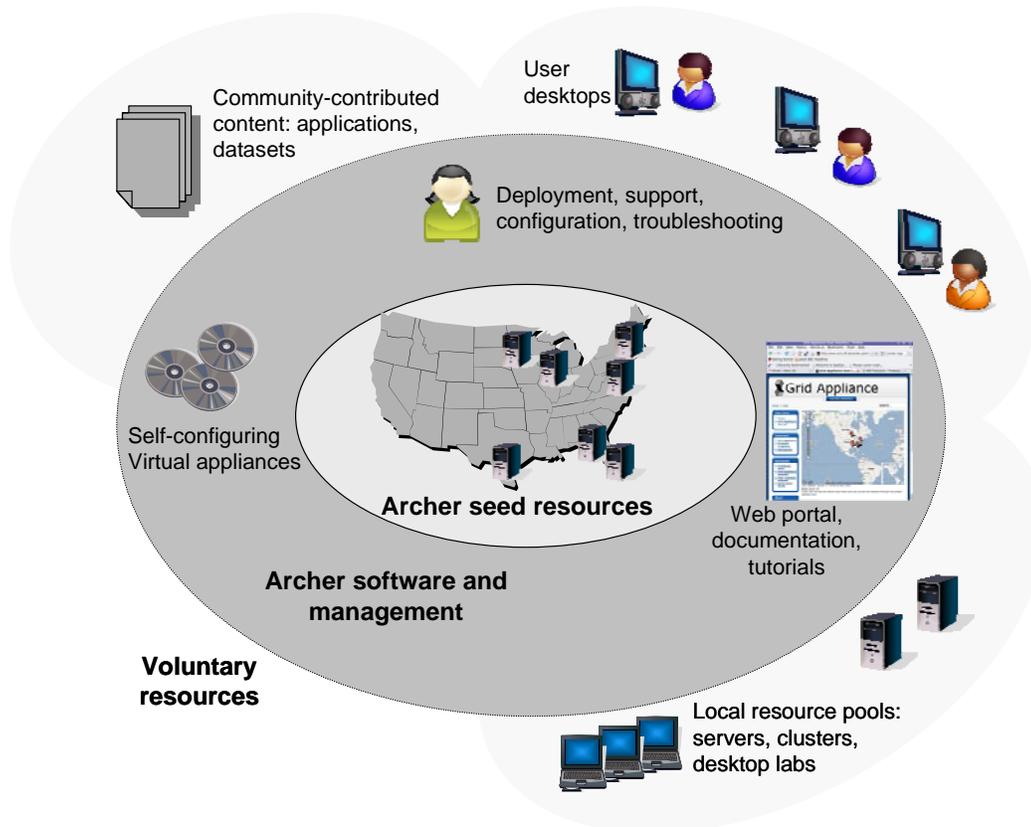

*Figure 1: Overview of Archer. The seed resources consist of seven clusters at Cornell, Florida State, Northeastern, Northwestern, U. Minnesota, U. Texas at Austin, and U. Florida. In addition to the hardware infrastructure, Archer provides virtual appliance and job scheduling software and ready access to user-contributed data and applications. Users from non-seed sites build upon these elements to increase available resources when they join the system.*

*Scenario 2: Local resource pooling and community sharing:* A group at Northeastern University has a local set of resources, time-shared and scheduled via ad-hoc scripts developed by students. Because the scripts do not provide load balancing, often resources become contended. They try out the Archer VM appliance and decide to join. Interacting with Archer management, they set up a local Condor pool. Their resources are load-balanced, and when not in use, they become available to other Archer users through Condor flocking.

*Scenario 3: Collaborative development and dissemination of tools and experiments*: A joint project between Cornell and Northwestern entails the development of an environment with extensions to the SESC simulator. Graduate students Carol at Cornell and John at NWU begin development by downloading code from the SESC software repository onto Archer appliances. Carol implements and tests new features in the simulator within her VM, creates Condor scripts that vary a parameter of interest, and places her code and scripts in a shared repository linked from the Archer Wiki. John uses Carol's code to perform experiments of his own. After several iterations, they gather data for their experiments and publish a paper highlighting their findings. They make the source code snapshot, benchmarks, and Condor scripts available on the Archer Wiki, enabling others to repeat and build upon their experiment.

*Table 1: Scenarios in which users with different levels of expertise can use Archer.*

| User | Use case scenario | Resources and interfaces used |
|---|---|---|
| Novice | Casual/trial usage of the system (e.g., homework assignments in undergraduate and graduate education). | Access pre-built tools, tutorials, educational modules through interactive Web portal. No local software required; only Web browser. |
| Entry-level | Undergraduate research; run small-scale experiments on Archer resources; graduate-level class projects. | Baseline Archer appliance installed on personal workstation. Entry-level user leverages existing simulation tools and job submit scripts. Advanced user builds simulation tools and scripts of their own. Software installation time: 15-30 minutes. |
| Advanced | Graduate research; run medium/large-scale experiments on Archer resources. Develop/modify simulation tools. | |
| Research groups | Use Archer software to manage local resources (e.g. desktop grids); deploy local/multi-site Archer pools with high priority for group users. | Customized Archer appliance installed on personal workstations of researchers, lab PCs, servers and clusters. Customization and installation times: hours to days. |

## 3 Archer Infrastructure

### 3.1 Overall design approach

The Archer infrastructure is a distributed system, motivated by scalability, sustainability and dependability arguments: new resources that join increase the system's capacity, the infrastructure is sustained by the community and does not overburden a single site with hosting, and the system can withstand hardware/software failures in individual sites. A distributed system, however, poses challenges in management which need to be addressed. Our system design builds on virtualization and autonomic computing techniques that specifically target ease of management. They make it possible to have effective centralized management of decentralized resources, similarly to successful infrastructures such as PlanetLab [16].

The Archer middleware integrates easy-to-install, self-configuring virtual machine appliances with virtual networks to create scalable community pools of virtual resources. Each Archer resource is a virtual appliance that is preconfigured with an installation of a Linux O/S and distributed computing middleware (Condor [15]). Archer virtual appliances are interconnected by the IPOP self-configuring virtual network overlay [10][11]. The choice of virtual appliances, virtual networks and Condor is motivated by the following reasons:

1. *Ease of deployment*: Virtual appliances can be easily deployed on typical x86-based machines regardless of their existing hardware/software configuration. Today's VM technologies are mature and several free virtualization options exist for Windows, Linux and MacOS systems (including VMware Player/Server, KVM, VirtualBox and Xen). Experiments with our prototype environment show that Archer virtual appliances can be deployed typically within 30 minutes by entry-level users.

2. *Software compatibility*: Virtual appliances can run unmodified, binary software, including a wealth of existing computer architecture simulators and support tools. Representative examples include SESC, SimpleScalar, PTLsim and Simics.

3. *Seamless connectivity*: The IPOP virtual network overlay which runs on Archer appliances provides bidirectional IP connectivity across all appliances. The virtual network supports nodes behind firewalls and network address translators (NATs) typical of educational institutions and Internet service providers. The virtual network is self-organizing and packaged with the virtual appliance in a way that does not require any configuration from end users.

4. *Scalable and robust job scheduling*: Condor is a robust job scheduler used in thousands of resources across the world. It supports both existing and Condor-linked applications, facilitates the queuing and management of large numbers of jobs, and has been successfully demonstrated to be effective in a variety of computer architecture studies.

5. *Isolation*: Virtual appliances are isolated from their hosts. Undesirable behavior is confined to a VM, which can be easily shut down and restarted from scratch by its user.

### 3.2 Archer core middleware

### 3.2.1 Virtual machines

Classic system VMs [20] were originally developed to enable efficient time-sharing of mainframe computers by multiple independent applications and O/Ss [13]. They are implemented by means of VM monitors (also known as hypervisors), which are responsible for intercepting and emulating the execution of privileged instructions that deal with shared resources: CPU, memory and I/O. VM technologies have evolved quite rapidly in recent years [17][9]. VMs now can achieve performance on par with non-virtualized systems [4], and are increasingly pervasive in commodity systems; virtualization extensions are shipped with all Intel and AMD x86 processors [26], virtualization software is available from a variety of vendors (VMware, Miscorsoft, Parallels) and in the open source realm (Xen [4]; KVM, which has already been integrated with the Linux kernel; and VirtualBox).

The isolation and decoupling properties of VMs are particularly attractive in distributed systems [1]. Virtual machines assist in the deployment of compute nodes because of their decoupling from the operating system running on the physical machine. VMs offer unique opportunities for load balancing and fault tolerance that build upon growing support for checkpointing and live migration of running VMs [7]. Furthermore, the ability to package VM software in easy-to-deploy virtual appliances [19] is attractive as a means to disseminate (and maintain) complex, preconfigured software and middleware stacks.

### 3.2.2 Virtual networks

Complementary to VMs, virtual networking enables isolated multiplexing of private networks providing the TCP/IP environment for communication among participating nodes [14][22][25]. Network virtualization techniques for distributed grid computing have been shown to provide applications their native network environments, despite the idiosyncrasies of the real physical network—in particular, the increasing use of Network Address Translation (NAT) and IP firewalls, recognized as a hindrance to programming and deploying distributed computing applications [21], does not impede the use of virtual network-based systems.

Archer VMs are decoupled not only from the physical hosts by means of the VM monitor, but also from the physical network by means of tunneling. Once instantiated, an Archer VM appliance is able to self-configure and maintain connections to other appliances via IPOP tunnels. The resulting system is akin in functionality to a Network of Workstations (NOW [2]); we term it a Wide-area Overlay of virtual Workstations (WOW) because both compute nodes and network links are virtualized, and resources are distributed across wide-area domains. Central to the scalability of WOWs are peer-to-peer discovery and routing techniques described in detail in [10][11].

### 3.2.3 Condor

Condor is an established distributed computing environment appropriate for building an ad-hoc HTC grid like Archer. The Condor Team has been engaged in constant research, software development and deployment of Condor for nearly 20 years [15]. Throughout this time, Condor has evolved from a local batch management system into a full-fledged distributed computing environment capable of supporting wide-area grids, complex workflows, compute-intensive applications, and data placement reliably, scalably, and with fault tolerance [24]. It has facilities for resource monitoring, job scheduling, and workflow supervision. Condor provides easy access to large amounts of dependable and reliable computational power over prolonged periods of time by effectively harnessing all available resources, including both dedicated compute clusters and non-dedicated machines under the control of interactive users or autonomous batch systems. Current statistics show that Condor has been deployed on well more than 100,000 computers in well more than 1400 Condor pools [28].

As part of the Grid Laboratory of Wisconsin (GLOW), Condor has built a local environment similar to the proposed Archer employing multiple Condor pools connected together to share resources between different research groups. Archer builds on this experience to create a similar grid across multiple institutions in a wide-area network. In particular, we leverage these features to deploy multiple Condor pools where priorities for remote and local users can be differentiated—local users can be in control of the policies that assign priorities and be able to configure higher-priority and preemptive scheduling to local users over remote users. This differentiation is an important feature of the Archer system, in that it creates an incentive for sites to join the infrastructure with several nodes. The pre-packaged Archer VMs provide an easy way to set up local Condor pools to manage jobs submitted by local users of a site, which are guaranteed to gain high-priority access to their resources and access to external Archer resources through flocking, while making their resources available to remote Archer users when they are idle. This kind of deployment with multiple shared pools where local control and priority is retained by individual groups has been an important feature of the GLOW infrastructure at Wisconsin, and we expect it to create further incentives for the growth of Archer.

### 3.3 Security considerations

By utilizing VMs, virtual networks and Condor, Archer provides several levels of isolation among users and with respect to the physical infrastructure. The only access that external users have to any Archer VM is through Condor, as an unprivileged user "nobody"—no direct logins are allowed. The VM runs only essential middleware services to minimize the possibility of privilege escalation within the VM. Even if privilege escalation does happen, users are confined to a virtual machine sandbox and do not have direct access to the underlying physical resources.

The TCP/IP traffic that is generated by a VM is completely confined to the virtual network, as described in [27]. Archer hosts are authenticated and traffic is encrypted end-to-end by deploying an security stack in each VM based on public key cryptography (PKI). In other words, Archer VMs are only able to communicate with other Archer VMs, preventing the use of Archer resources to initiate denial of service or other kinds of attacks to physical resources. There are a couple of exceptions to this rule, which are necessary for Archer VMs to be accessible from physical resources so users can interact with them. We establish communication channels between Archer VMs and physical hosts using host-only virtual networks (software-emulated networks confined to a single host) that are carefully controlled to provide only two types of services: secure shell logins, and access to user data within the VM through Samba and NFS file systems.

In a typical usage case of a Windows-based desktop, a user deploys an Archer VM on their desktop, interacts with the X11-based graphical user interface in the VM through its console, logs into the VM from the physical host using SSH, and browses a Samba network share exported by the appliance (accessible only within the host) to copy data to and from the VM.

Security patches are regularly applied to the baseline Archer VMs and made available for upgrades; the process of upgrading VM appliances is facilitated by the use of UnionFS stacked file systems, whereby it is possible for users to upgrade the base system configuration of the appliance by simply replacing a virtual disk file and rebooting the system. User data and local configurations remain unmodified in the VM after the upgrade as they are stored in different stacks of the file system [27].

### 3.4 Performance considerations

The advantages in isolation, security and management provided by VMs connected over virtual networks have associated performance overheads due to the VM and virtual network tunneling. However, these overheads are often small for CPU-intensive applications which are typical in computer architecture simulation. Studies have quantified this overhead under different scenarios, showing that the overhead CPU-intensive applications such as SPEC benchmarks is a few percent points [4][8], and in [27] it has been shown that the virtual/physical overhead for a Xen VM connected to a virtual network approximately 1% for a 37-minute Condor-scheduled SimpleScalar run (sim-cache, "go" benchmark). The performance of Condor in VM environments is also studied in [18].

The following experiment illustrates the capabilities and performance of the middleware and software of Archer in the context of computer architecture research and education. We have run a simulation experiment in our prototype Archer deployment, in which 200 jobs were submitted from a virtual appliance. Each job consisted of a cache simulation running 1 billion instructions of the SPEC benchmark "go" for different cache configurations arising from varying level-1 and level-2 cache sizes and associativities. The jobs were submitted from a laptop running the Archer virtual appliance behind a broadband (1MB/s) network provider. The virtual appliances in which the jobs executed were distributed across five universities (including U. Florida, U. Minnesota, Northwestern University), making up a pool of 56 VMs.

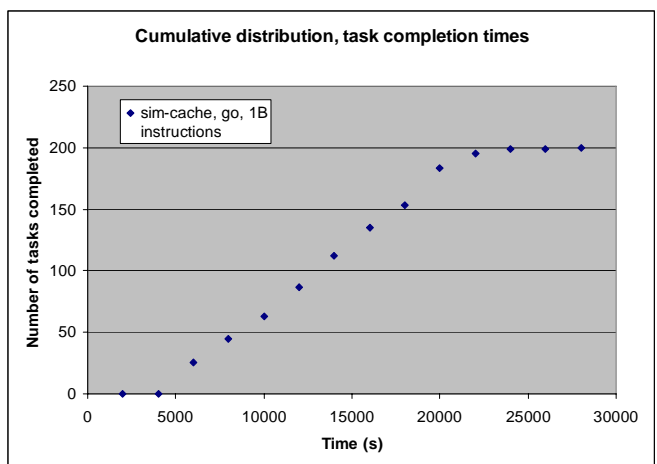

*Figure 2: Distribution of simulation completion times for an experiment with 200 SimpleScalar sim-cache jobs executed on a 56-node prototype Archer system. The median and average single job execution times are 4080 and 4320 seconds, respectively. In steady state the system was completing jobs at an approximate rate of one job every 90 seconds, compared to the throughput of one job per 42 minutes of a single job running on a single resource.*

The total execution time to finish all 200 jobs was approximately 7.5 hours. If these jobs were to be executed on a single node, the execution time would have been 9.5 days assuming the median single job times measured across the 56 heterogeneous resources. Figure 2 shows a plot of the cumulative distribution of number of jobs completed as a function of time. The virtualization overhead for this application using VMware-based VMs was measured to be 11 percent, which is acceptable given the goal of achieving high throughput. Reducing the overhead introduced by virtual machines is an active area of research and development in academia and industry, and the expected trend is for these overheads to be reduced. With the use of a different VM technology (Xen) which we also plan to support on Archer, the overhead due to virtualization for this simulator is only 1% [27].

**4 Related Work**

Inspired by projects which have been extremely successful at bringing large number of voluntary resources, such as SETI@home [3] and other BOINC-based (Berkeley Open Infrastructure for Network Computing) projects, Archer also allows nodes to join the infrastructure seamlessly. The key difference in Archer is that, in BOINC-based systems, applications need to be modified to use their application programming interfaces, and users are constrained to donating resources only. In contrast, in the Archer infrastructure, the computing node sandboxes are system VMs capable of running existing, unmodified binary applications, which is critical for adoption by the computer architecture community. Furthermore, users are able to *both* donate and make use of Archer resources through their virtual appliances.

Archer is closely related to PlanetLab [16] (www.planet-lab.org) with respect to how resources are distributed and managed. PlanetLab is also a distributed system where individual researchers across many sites contribute to the overall aggregate capacity of the system by providing locally managed physical hardware (805 nodes at 402 sites worldwide, as of July 2007, while the middleware and software is managed in a centralized manner (by PlanetLab Central). However, Archer differs fundamentally from PlanetLab in purpose. PlanetLab is a generic testbed for experimental networking research and does not support load balancing of jobs, while Archer targets compute-intensive applications. Archer is also different in that it does not require dedicated non-firewalled physical machines.

RAMP (Research Accelerator for Multiple Processors, ramp.eece.berkeley.edu) is a related resource for the computer architecture community. The focus of RAMP is on the use of programmable logic to speed up the simulation of large-scale multiprocessors. While RAMP provides the potential for large speedups over software simulation, it requires users to develop their simulation infrastructure to match the specific RAMP software and hardware stack. Archer, in contrast, is general-purpose and supports a wealth of unmodified single- and multi-processor simulation tools that computer architecture researchers already use in their own local environments (e.g. SimpleScalar [5], SESC [29]), offering a lower barrier of entry to its use. Nonetheless, Archer and RAMP are complementary resources in that they focus on different aspects of quantitative computer architecture research: general-purpose simulation in Archer, high-performance multi-processor simulation in RAMP.

Archer is related to OSG (www.opensciencegrid.org), where resources are pooled across institutions using a consistent software base packaged for ease of configuration, deployment and maintenance of middleware (Virtual Data Toolkit, VDT), and TeraGrid (www.teragrid.org), a high-performance infrastructure well-suited to run large parallel jobs. Aside from the fundamental differences in goals outlined in Section C.1, Archer differentiates from these systems with respect to its technology: the use of VM-based appliances for software distribution and self-configuring virtual networking to facilitate the addition of resources to the infrastructure. Also, Archer is targeted at serving a single rather than multiple communities, which enables its content to be tailored to the interest of computer architects by the architecture community.

Archer is similar to Intel's NetBatch infrastructure in its support for high-throughput simulation workloads. NetBatch is also a distributed system consisting of CPUs distributed across multiple sites, managed by an in-house batch scheduler [6]. It has been highly successful in providing batch computing cycles for a variety of applications at Intel: it started with hundreds of computers in 1990, and over the course of ten years grew to 10,000 nodes across 25 sites, logging 2.7 million jobs per month in their queues [12]. Archer is different from NetBatch in that it is not internal to a private corporate network, allowing individuals to easily join and contribute resources to the system.

## Acknowledgments

Archer is supported by the National Science Foundation under CRI Grants 0751112, 0750847, 0750851, 0750852, 0750860, 0750868, 0750884, and 0751091. Any opinions, findings and conclusions or recommendations expressed in this material are those of the authors and do not necessarily reflect the views of the National Science Foundation.